\documentclass[USenglish,twocolumn]{article}

\usepackage[utf8]{inputenc}%(only for the pdftex engine)
\usepackage[big]{dgruyter}
  
\begin{document}

\articletype{Research Article{\hfill}Open Access}

\author*[1]{St\'ephane Vennes}
\author[2]{P\'eter N\'emeth}
\author[3]{Adela Kawka}

\affil[1]{Astronomical Institute of the Czech Academy of Sciences, CZ-251\,65, Ond\v{r}ejov, CR, E-mail: vennes@asu.cas.cz}
\affil[2]{Astroserver.org, 8533 Malomsok, Hungary, E-mail: peter.nemeth@astroserver.org}
\affil[3]{Astronomical Institute of the Czech Academy of Sciences, CZ-251\,65, Ond\v{r}ejov, CR; E-mail: kawka@asu.cas.cz}

\title{\huge A FEROS Survey of Hot Subdwarf Stars}
\runningtitle{S. Vennes et al., Hot Subdwarf Stars}

%\subtitle{...}

\begin{abstract}
{We have completed a survey of twenty-two ultraviolet-selected hot subdwarfs using the
Fiber-fed Extended Range Optical Spectrograph (FEROS) and the 2.2-m
telescope at La Silla. 
The sample includes apparently single objects as well as
hot subdwarfs paired with a bright, unresolved companion. The sample was extracted from
our $GALEX$ catalogue of hot subdwarf stars. We identified three new short-period systems ($P=3.5$ hours to 5 days) and determined the
orbital parameters of a long-period ($P=62^{\rm d}.66$)
sdO plus G III system. This particular system should evolve into a close double degenerate system
following 
a second common envelope phase. We also conducted a chemical abundance study of the subdwarfs: Some objects show nitrogen and argon abundance excess with respect to oxygen. We present key results of this programme.}
\end{abstract}

%  \classification[PACS]{}
% \communicated{...}
% \dedication{...}

\journalname{Open Astronomy}
\DOI{DOI}
  \startpage{1}
  \received{..}
  \revised{..}
  \accepted{..}

  \journalyear{2017}
  \journalvolume{1}
%  \journalissue{1}
 
\maketitle

\section{Introduction}

The properties of extreme horizontal branch (EHB) stars, i.e., the
hot, hydrogen-rich (sdB) and helium-rich subdwarf (sdO) stars, located at the faint blue end of the
horizontal branch (HB) are dictated by binary interaction processes arising at critical stages of their evolution \citep{men1976}. Hot subdwarf stars are compact, helium core-burning objects with high effective temperatures ($\gtrsim 20\,000$~K) but with a luminosity comparable to that of cooler HB stars.

\citet{dor1993} reproduced the properties of hot subdwarf stars with evolutionary models characterized by extremely thin hydrogen envelopes ($\lesssim 0.001\,M_\odot$). The removal of the hydrogen envelope must occur while the primary ascends the red giant branch and, depending on the binary mass ratio, involves unstable mass transfer via a common envelope (CE) or stable mass transfer via a Roche lobe overflow (RLOF). The existence of single hot subdwarfs may be ascribed to merger events \citep{web1984,han2002}. 

Based on population syntheses, \citet{han2003} find that a majority (up to 90 percent) of subdwarfs form during a mass transfer event, i.e., comparable to the observed fraction ($\approx$70\%) of short period binaries \citep[$P<10$~d, ][]{max2001,mor2003}, while the remaining single objects are formed by the merger of two helium white dwarfs which would also result in a thin hydrogen layer. \citet{han2003} predict a binary period distribution ranging from 0.5~hr to 500~d, or even 1600~d according to \citet{che2013}.

Our knowledge of the binary fraction and period distribution is based on a sample of nearly 200 known systems \citep[see ][]{kup2015,kaw2015} which may be affected by selection effects. Radial velocity surveys such as the ESO Supernovae type Ia Progenitor surveY \citep[SPY, ][]{nap2004} or surveys based on the Palomar-Green and Edinburgh-Cape catalogues \citep{cop2011}, and the SDSS \citep{gei2011,kup2015} or $GALEX$ samples \citep{kaw2015} have confirmed the prevalence of short-period binaries but at a lower fraction than predicted. 

Overall,
the peak of the distribution is close to 1~d and most systems have a period between 2~hr and 10~d. A few systems
have been identified with a period between 1 and 3 yr \citep{bar2013,dec2017,vos2017,vos2018}, but very few with a period between 30 and 300~d. Long-period systems with a large mass ratio (F- or G-type companion) emerge from a RLOF that result in increased binary separations, while short-period, low mass-ratio systems (M dwarf or WD companion) emerge from a CE phase. The secondary mass distribution shows two main peaks, one at $\approx 0.1~M_\odot$ and populated with late-type dwarfs and another one close to $0.5~M_\odot$ and dominated by white dwarfs. High-mass white dwarfs are also evident in the distribution. 

In the following sections we describe a high-dispersion survey of ultraviolet-selected, hot subdwarf stars and highlight some key results of our survey. First, we report the discovery of three new close binaries and, next we describe the properties of an evolved binary comprised of an sdO primary star paired with a red dwarf companion. Finally, we present selected results of a new abundance study for a few members of the sample.

\begin{table*}[!]
\begin{center}
\caption{Hot subdwarf sample: identification and class.\label{tbl_targets}}
\begin{tabular}{llcccc}
\hline
GALEX & Name & $V$ (mag) & Spectral Type & Companion class & Notes\\
\hline
J011525.92$+$192249.9 &                               & 13.09 & sdB+F2V & MS & \\
J011627.22$+$060314.2 & PB 6355                       & 13.24 & sdB+F6V & MS & \\
J021021.86$+$083058.9 &                               & 13.40 & sdB+F2IV & SG & \\
J022454.87$+$010938.8 &                               & 12.32 & sdB+F4V & MS & \\
J025023.70$-$040611.0 & TYC 4703-810-1, HE 0247-0418  & 13.02 & sdB & WD: & a \\
J040105.31$-$322346.0 & EC 03591-3232, CD-32 1567     & 11.20 & sdB & --- &\\
J050720.16$-$280224.8 & CD-28 1974, EC 05053-2806     & 12.39 & sdB+G & MS & \\
J081233.60$+$160121.0 & SDSS J081233.67+160123.7      & 13.57 & sdB & WD: & a \\
J082832.80$+$145205.0 & TYC 808-490-1                 & 11.65 & He-sdB & --- & \\
J085649.30$+$170115.0 & LAMOST J085649.26+170114.6    & 13.17 & sdB & --- & \\
J093448.20$-$251248.0 & TYC 6605-1962-1               & 13.03 & sdB & RD: & a\\
J095256.60$-$371940.0 & TYC 7180-740-1                & 12.69 & He-sdO &  --- & \\
J135629.20$-$493403.0 & CD-48 8608, TYC 8271-627-1    & 12.30 & sdB & --- & \\
J163201.40$+$075940.0 & PG 1629+081                   & 12.76 & sdB &  WD & b\\
J173153.70$+$064706.0 &                               & 13.74 & sdB & WD & b \\
J173651.20$+$280635.0 & TYC 2084-448-1                & 11.44 & sdB+F7V & MS &  \\
J175340.57$-$500741.8 &                               & 12.88 & sdB+F7V & MS & \\
J203850.22$-$265750.0 & EC 20358-2708, TYC 6916-251-1 & 11.90 & sdO+G3III & RG & a \\
J220551.86$-$314105.5 & TYC 7489-686-1                & 12.41 & sdB & MS (RD) & b\\
J222758.59$+$200623.3 & TYC 1703-394-1                & 10.60 & sdB+F5V & MS & \\
J225444.11$-$551505.6 & TYC 8827-750-1                & 12.08 & sdB & WD & b\\
J234421.60$-$342655.0 & MCT 2341-3443, CD-35 15910    & 11.39 & sdB & --- &  \\
\hline 
\end{tabular}\\

Notes: (a) new radial velocity variable star; (b) known radial velocity variable star \citep[see ][]{kaw2015}.
\end{center}
\end{table*}

\section{MPG2.2m/FEROS survey}

We conducted the survey using FEROS (The Fiber-fed Extended Range Optical Spectrograph) attached to the MPG 2.2-m telescope at La Silla. The spectrograph offers a resolution $R=48\,000$ and covers a nominal spectral range from 3\,500 to 9\,000\AA.
Some 136 spectra were obtained during six observing runs between November 2014 and September 2017.
The targets were selected from a UV-based catalogue of hot subdwarf stars prepared by \citet{ven2011} and \citet{nem2012}. All systems are bright photometric sources ($NUV<14$~mag) in the $GALEX$ all-sky survey \citep{mor2007}. The spectroscopic identifications were secured with low-dispersion spectra. Table~\ref{tbl_targets} lists members of the present sample. The sample comprises 22 objects including four close binary systems previously studied by \citet{kaw2015}. Half of the sample of new objects are systems with a bright companion star, while the other half are spectroscopically single subdwarfs. We show that three out of these nine apparently single objects are in fact paired with an unseen companion.

\section{Analysis}

The systems with significant radial velocity variations may be segregated in two broad classes: those with unseen companions, i.e., a red dwarf (dM) or a white dwarf,
 and those with spectroscopically identifiable companions, i.e., F, G, or K-type stars.
White dwarf companions are suspected in systems with a relatively large mass function but without phased reflection effects from a close red dwarf companion. Several, apparently single hot subdwarfs do not show radial velocity variations including J0401-3236, the lead-rich hot subdwarf J0828+1452 \citep{jef2017}, J0856+1701, the He-sdO J0952-3719, J1356-4934, and J2344-3426.

The survey also included known composite binaries \citep[see ][]{nem2012} with spectra suitable for decomposition and radial velocity measurements. Details of this study will be presented elsewhere (Vennes, Nemeth \& Kawka, 2018, in preparation).

We performed a preliminary abundance analysis of suitable spectra with a high signal-to-noise ratio. The model atmosphere and
spectral synthesis were computed using {\sc Tlusty/Synspec} \citep{hub1995,lan1995,hub2017} and the spectra were analyzed using the multi-parameters fitting procedure {\sc XTgrid} \citep{nem2012}. The atmospheres are computed in non-local thermodynamic equilibrium (non-LTE) and include all relevant elements up to iron (H, He, C, N, O, Mg, Al, Si, S, Ar, Fe).

\subsection{Subdwarf plus white dwarf}

\begin{figure}
\resizebox{8.3cm}{!}{\includegraphics[viewport=30 285 570 560, clip]{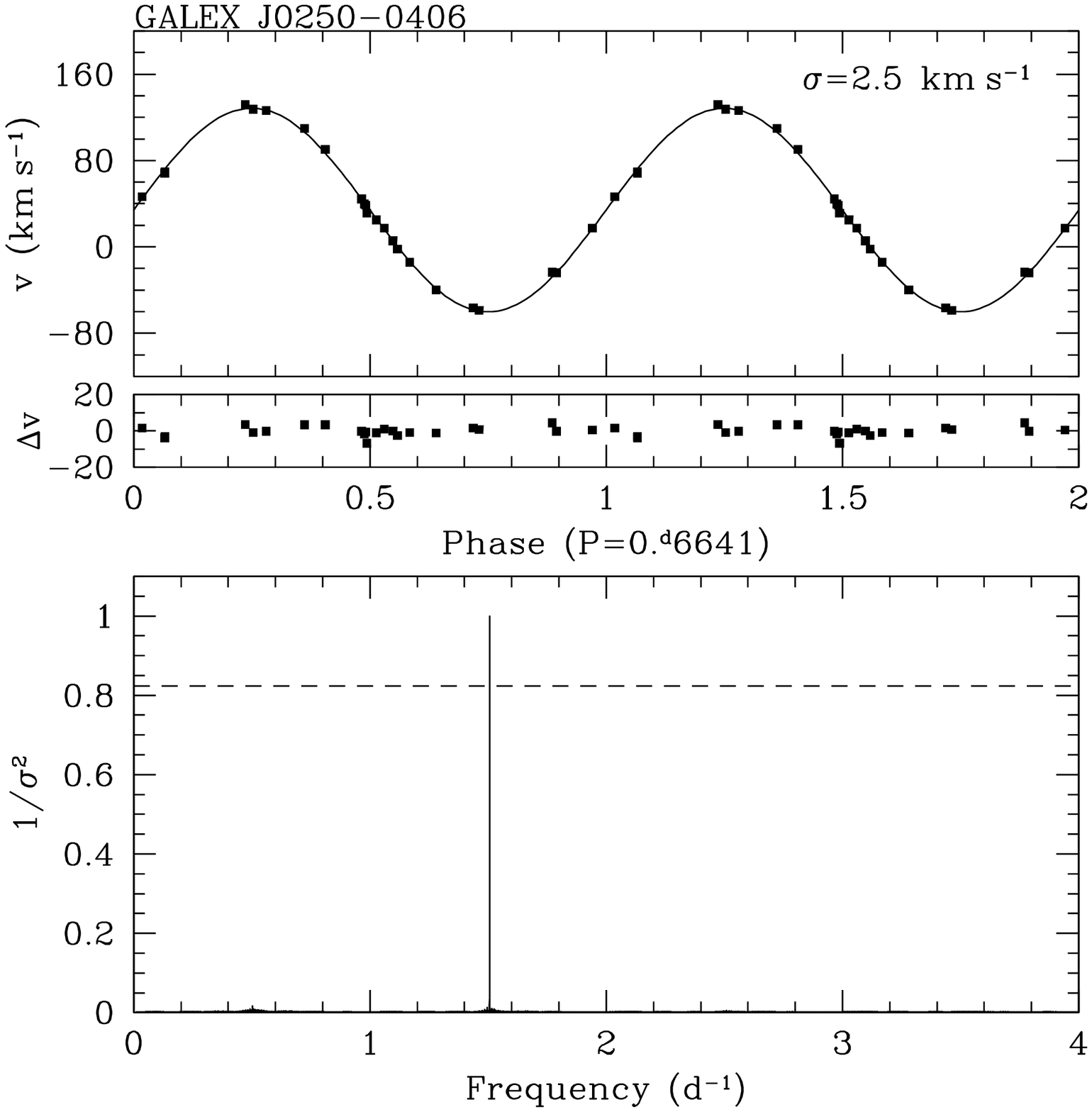}}
\caption{Radial velocity measurements of the sdB star folded on the orbital period ($P=15.9$ hr).
The NSVS light-curve
does not show variability when folded on the orbital period which
suggests the likely presence of a normal white dwarf companion.\label{fig1}}
\end{figure}

The hot subdwarf J0250-0406 ($T_{\rm eff}=28\,560$\,K, $\log{g}=5.67$) is in a close binary with a suspected
white dwarf companion (Fig~\ref{fig1}). The mass function 
$f=0.057\pm0.03\,M_\odot$ implies a minimum secondary mass of $0.33\,M_\odot$ assuming a subdwarf mass of $0.47\,M_\odot$.
The NSVS \citep[Northern Sky Variability Survey, ][]{woz2004} photometric time series do not show variability with a semi-amplitude $\lesssim 15$~mmag when phased on the orbital period ($P=0.6641$~d) which rules out the presence of a close red dwarf companion since a phase-dependent reflection effect would reach a semi-amplitude of $\approx$0.2~mag \citep{max2004}. If indeed the secondary is a normal white dwarf then a mass of $0.6\,M_\odot$ is obtained at an inclination of $42^\circ$.

The hot subdwarf J0812+1601 ($T_{\rm eff}=31\,580$\,K, $\log{g}=5.56$) is in a longer period binary with a high mass function ($f=0.07\pm0.02\,M_\odot$) that implies a minimum mass
of $0.37\,M_\odot$ for the companion also assuming a subdwarf mass of $0.47\,M_\odot$. The NSVS time series do not show variability on an orbital period of $5.1$ day with a semi-amplitude $\lesssim 30$~mmag, i.e., much lower than the expected reflection effect of 0.12~mag semi-amplitude if the companion is a red dwarf. We
conclude that the companion is most likely a white dwarf.
Improved orbital parameters for several close subdwarf plus white dwarf binaries studied by \citet{kaw2015}---J1632+0759, J1731+0647, and J2254-5515--- will be presented elsewhere (Vennes, Nemeth \& Kawka, 2018, in preparation). 

\subsection{Subdwarf plus red dwarf}

\begin{figure}
\resizebox{8.3cm}{!}{\includegraphics[viewport=30 285 570 560, clip]{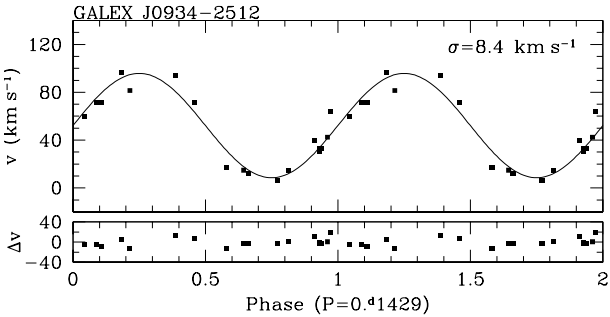}}
\caption{Radial velocity measurements of the sdB star folded on the orbital period ($P=3.43$ hr). A SuperWASP light-curve does not show any variability when folded on this period, but
the mass function implies a very low secondary mass, or, less probably that the companion is a white dwarf and system is seen at a low inclination.\label{fig2}}
\end{figure}

The hot subdwarf J0934-2512 ($T_{\rm eff}=34\,440$\,K, $\log{g}=5.17$) is in a very short period binary ($P=3.43$ h, Fig.~\ref{fig2}) and has a low mass function ($f=0.0011\pm0.0004$) which lead us to suspect the presence of a close red-dwarf companion ($M>0.07\,M_\odot$). However, the SuperWASP \citep[Wide Angle Search for Planets, ][]{pol2006} photometric time series do not show variability with a semi-amplitude $\lesssim10$~mmag. However, at 0.07$\,M_\odot$, the radius of the red dwarf is only 0.11$\,R_\odot$ and the semi-amplitude of the phase-dependent variations is $\approx$17~mmag which is marginally consistent with the measured upper limit. A lower inclination implies a larger secondary star which increases reflected light.
On the other hand, the presence of a $>0.5\,M_\odot$ white dwarf would only be possible at an improbably (<2\%) low inclination of $<12^\circ$. We conclude that the secondary is most likely a low-mass red dwarf.

This radial velocity survey was also an opportunity to finalize our study of the close subdwarf plus red dwarf binary J2205-3141 \citep{kaw2015} and demonstrate strict phasing of the reflection effect with the orbital ephemeris. A detailed abundance study based on this new data set is shown in a following section.

\subsection{Subdwarf plus red giant}

\begin{figure}
\resizebox{8.3cm}{!}{\includegraphics[viewport=30 285 570 560, clip]{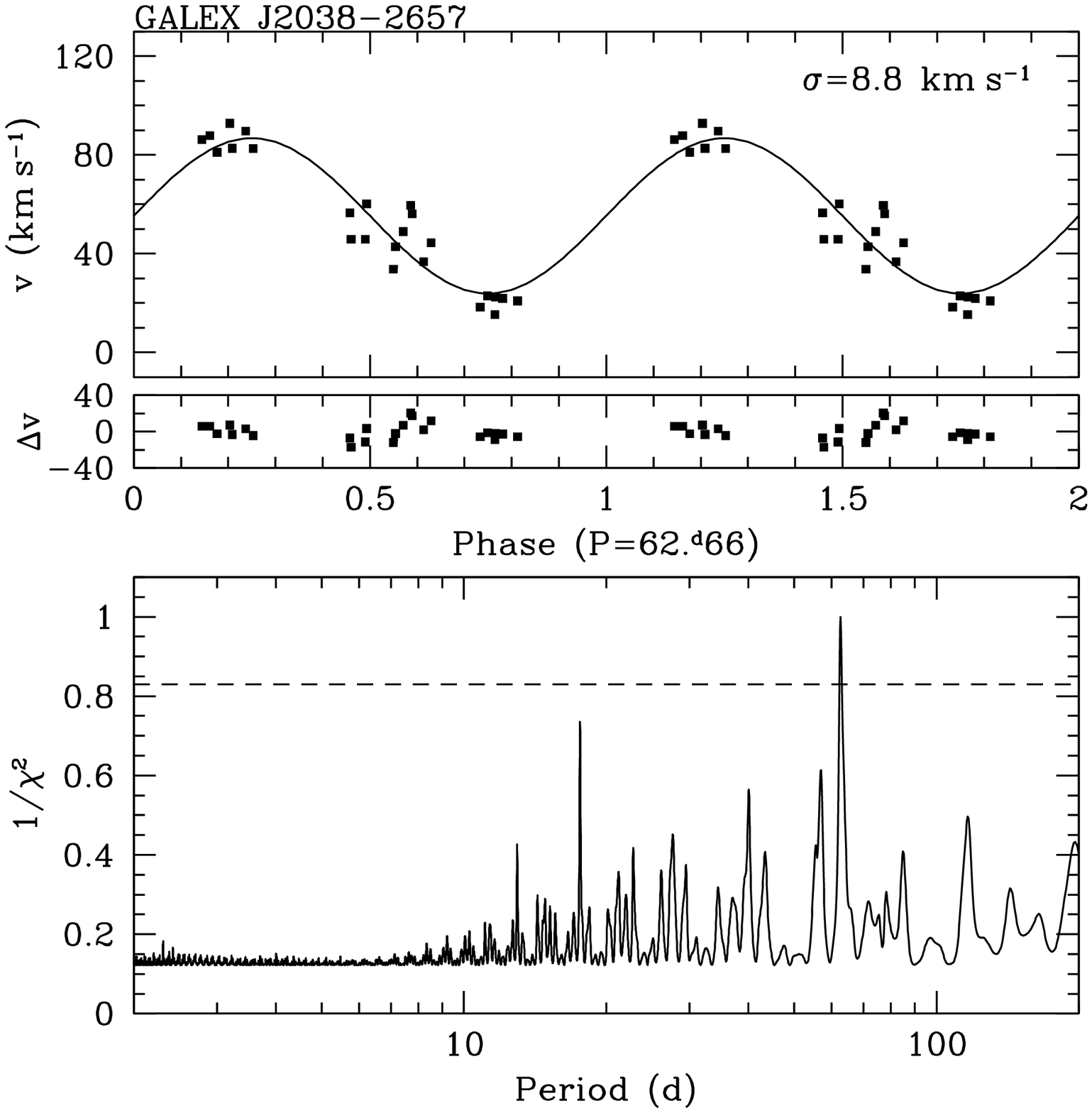}}
\caption{Radial velocity measurements of the sdO star folded on the orbital period ($P=62.66$ days). The SuperWASP light-curve shows double-peaked, short-period variations \citep{kaw2015}.\label{fig3}}
\end{figure}

The binary GALEX~J2038-2657 \citep[$=$EC20358-2708, ][]{odo2013} is at a critical evolutionary stage sitting in between two consecutive CE phases.
Fig~\ref{fig3} shows the orbital analysis based on
our radial velocity survey conducted at La Silla.
The hot subdwarf ($T_{\rm eff}=58\,450$\,K, $\log{g}=5.04$) remains subjugated by its companion across the optical range.
The bright companion star is already
sitting on the giant branch and
will eventually climb the asymptotic giant branch (AGB) and initiate the second CE phase leading to the
formation of a close double degenerate system and a potential pre-SNIa merger event.

\begin{figure}[!t]
\resizebox{7.5cm}{!}{\includegraphics[viewport=0 20 540 290, clip]{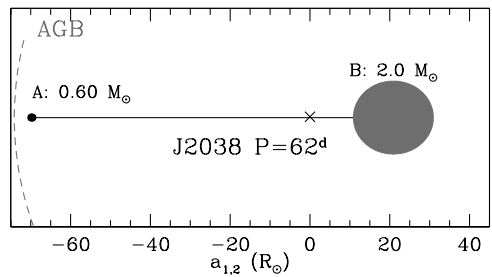}}
\caption{Schematic representation of the present-day sdO+GIII binary GALEX~J2038-2657 also showing the projected radius of the future
AGB star. The sdO star should enter a CE phase with the AGB star, and, by that time, the hot sdO star may already have settled on the white dwarf cooling track. \label{fig4}}
\end{figure}

The most likely evolutionary scenario begins with two main sequence stars in a long period binary ($>>$100 d) with a higher-mass ($>>$2 M$_\odot$) progenitor
for the present-day hot sdO star (star A) and a lower-mass ($\approx$2M$_\odot$) progenitor of the present-day red giant (star B). The system entered
the first CE phase after star A climbed on the red giant branch dispersing the envelope and settling on the extreme horizontal branch with a period of 62.66~d. 
Fig.~\ref{fig4} illustrates the present day binary with a separation of 90$R_\odot$. When star B eventually climbs the AGB it will engulf its companion leading to a second CE phase and
dramatic orbital shrinkage. At this stage, star A may have exhausted its nuclear supply and initiated its descent onto a typical $0.6-0.7\,M_\odot$ white dwarf cooling track.  Almost concurrently, star B will terminate its nuclear life and initiate its own descent onto a $0.6-0.7\,M_\odot$ white dwarf cooling track. The end result will be a close 1.2-1.4$\,M_\odot$ double degenerate star which in due time should be considered as a viable Type Ia supernova progenitor.\\

\begin{figure*}
\begin{center}
\resizebox{14cm}{!}{\includegraphics[viewport=50 45 395 565, clip]{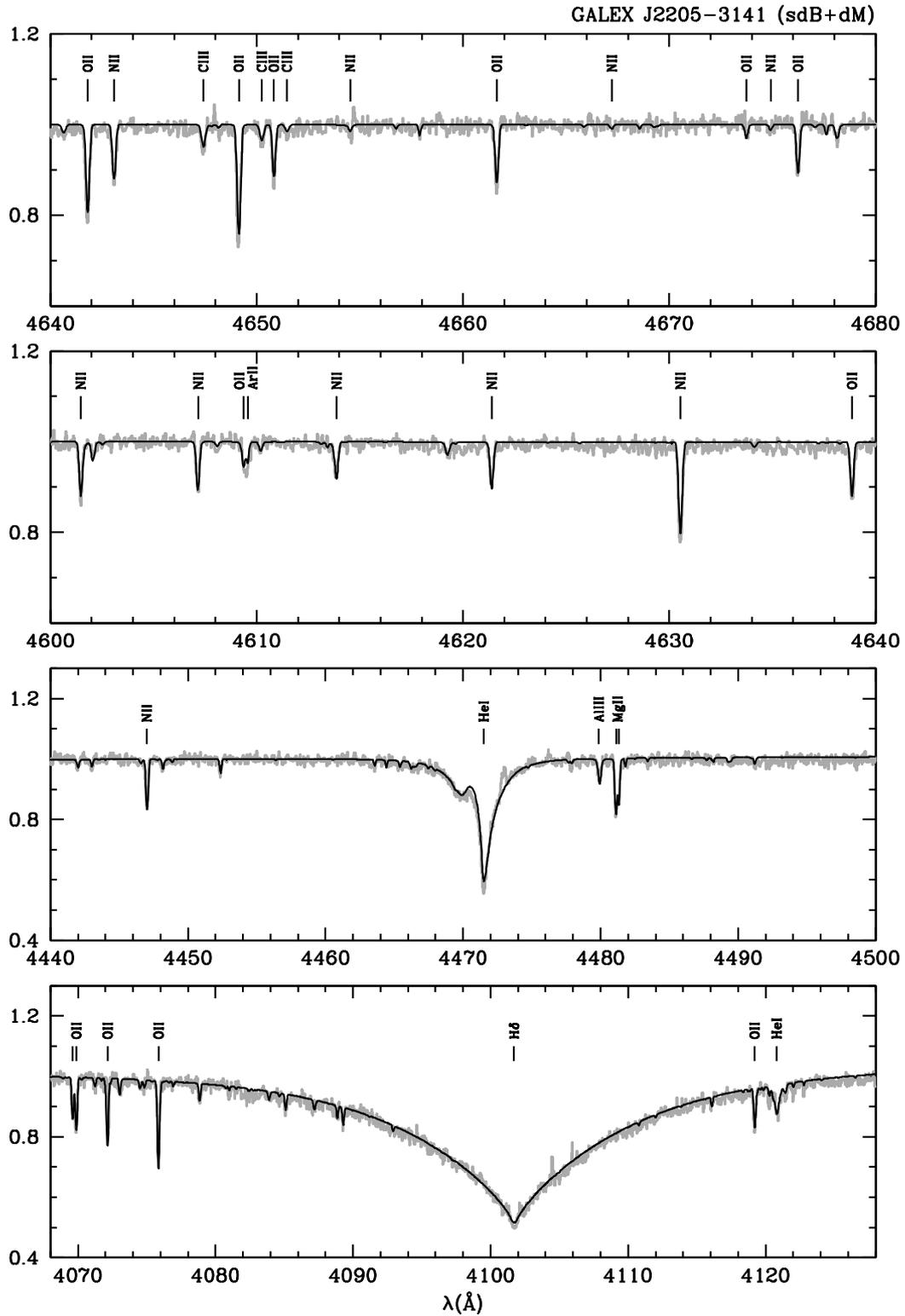}}
\caption{High-dispersion spectra of the hot subdwarf J2205-3141 obtained
with FEROS and best-fit non-LTE model (Tlusty/Synspec). The object is in a short period (8.2 hr) binary with a late-type companion
showing a strong reflection effect \citep{kaw2015}. The hot subdwarf exhibits near-solar abundance for most
elements and a remarkable excess of nitrogen and argon with respect to oxygen.\label{fig5}}
\end{center}
\end{figure*}

\subsection{Abundance study}

Abundance analyses are performed using the co-added and high signal-to-noise ratio spectra. Doppler corrections are applied to individual spectra which are then set to the rest frame. Fig~\ref{fig5} shows the co-added FEROS spectrum of the close binary J2205-3141 in four segments showing numerous spectral lines suitable for an abundance analysis. The models and fitting techniques are described in \citet{nem2012}. 

\begin{figure}[!t]
\resizebox{8.9cm}{!}{\includegraphics[viewport=0 0 580 580, clip]{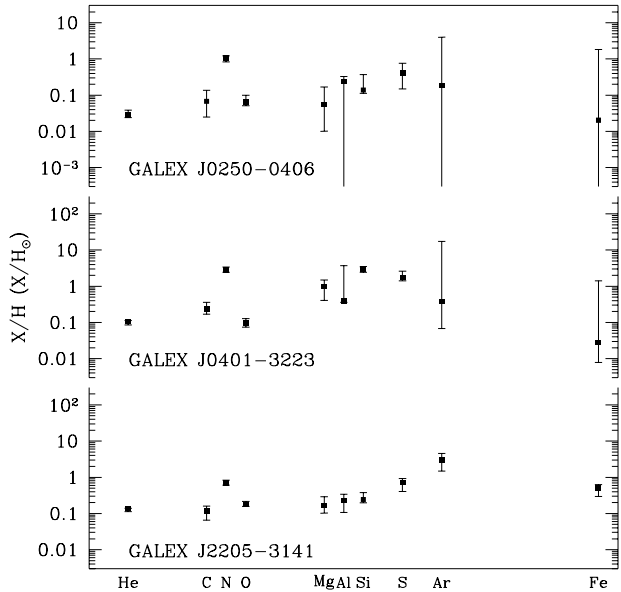}}
\caption{Abundance measurements relative to the Sun in three hot hydrogen-rich subdwarfs: J2205-3141
comprised of an sdB primary and a late-type companion, J0250-0405 comprised of an sdB and
a likely white dwarf companion, and J0401-3223 which is possibly single or in a very long period system. All three hot subdwarfs exhibit nitrogen excess relative to oxygen and other elements.\label{fig6}}
\end{figure}

Fig~\ref{fig6} shows the measured abundance patterns for three hot subdwarfs, J0250-0406 which is in a close binary with a white dwarf star, J0401-3223 which is apparently single or in a wide binary, and J2205-3141 which is in a close reflection binary with a red dwarf. The abundance patterns are compared to solar abundances. Light element abundances already show an interesting peculiarity: In all three systems, the nitrogen abundance exceeds that of carbon and in two cases even that of oxygen ([N$/$O]$>0.5$ dex). Intermediate elements follow near solar pattern with the exception of argon which is overabundant relative to oxygen in the atmosphere of J2205-3141 ([Ar$/$O]$\approx 1.0$ dex). The complete abundance pattern will eventually include all of the iron-group and heavier elements. 

Evolutionary effects may alter the surface composition of some hot subdwarfs \citep[see, e.g., ][]{swe2004}. However, atomic diffusion in hot subdwarfs may also lead to peculiarities that erase evolutionary effects \citep{ung2001}. Such induced peculiarities are directed by present day stellar parameters ($T_{\rm eff}, \log{g}$). Vertical abundance inhomogeneities are also expected in the atmosphere leading to line profile distortions relative to homogeneous atmospheres. 

\section{Summary}

We have completed a spectroscopic survey of hot subdwarf stars using the MPG2.2-m/FEROS at La Silla.
We have identified three new evolved binaries comprised of a hot subdwarf primary star and, in two cases, a likely
white dwarf companion, or a red dwarf companion in the remaining case. We have also completed a radial velocity study of
the hot subdwarf plus red giant binary J2038-2657 and concluded that the system will likely enter another CE phase while
the secondary star climbs the AGB leading to the formation of a close double degenerate system. 
High signal-to-noise ratio echelle spectra are also suitable for
abundance studies \citep[see ][]{gei2013}. We presented sample results for three hot subdwarfs: J0250-0406 which resides in a close binary
with a white dwarf, J0401-3223 which is apparently single, and J2205-3141 which is paired with a red dwarf companion.
The atmosphere of all three objects shows a nitrogen overabundance relative to oxygen, while the hot subdwarf J2205-3141 also shows an excess of argon relative to oxygen. 

\begin{acknowledgement}

AK and SV acknowledge support from the Grant Agency of the
Czech Republic (15-15943S) and the Ministry
of Education of the Czech Republic (LG14013). We would like to thank L. Zychova, M. Piecka, P. Kabath, and S. Ehlerova for their
assistance with the observations.
This research made use of services at Astroserver.org under reference code
VPD0AJ.
\end{acknowledgement}

{} 

\end{document}